\documentclass[12pt,a4paper]{article}
\usepackage[utf8x]{inputenc}
\usepackage[T1]{fontenc}
\usepackage{amsfonts}
\usepackage{amsmath}
\usepackage[pdftex]{graphicx} 
\usepackage[pdftex,linkcolor=black,pdfborder={0 0 0}]{hyperref} 
\usepackage{calc} 
\usepackage{enumitem} 
\usepackage{amsthm}
\usepackage{bm}
\usepackage{xcolor}
\usepackage{natbib}
\usepackage{placeins}
\usepackage{comment}
\usepackage{algorithm2e}
\RestyleAlgo{ruled}
\usepackage{dsfont}

\def\spacingset#1{\renewcommand{\baselinestretch}%
{#1}\small\normalsize} \spacingset{1}

\DeclareMathOperator{\UD}{U}
\DeclareMathOperator{\ND}{\mathcal N}

\DeclareMathOperator{\BetaD}{Beta}
\newcommand\norm[1]{\lVert#1\rVert}
\newcommand\indicator[1]{\mathds{1}_{\left\{ #1 \right\}}}
\DeclareMathOperator{\Nu}{\mathcal{V}}

\begin{document} 

\begin{titlepage}

\title{Calibrated Multivariate Regression with Localized PIT Mappings}

\author{Lucas Kock$\mbox{}^1$, G.~S.~Rodrigues$\mbox{}^{2*}$, Scott A.~Sisson$\mbox{}^3$,\\ Nadja Klein$\mbox{}^4$ and David J. Nott$\mbox{}^1$} 

\date{\today}
\maketitle
\thispagestyle{empty}
\noindent
\vspace{2em}

\begin{center}
{\Large Abstract}
\end{center}
\vspace{-1pt}
\noindent Calibration ensures that predicted uncertainties align with observed uncertainties. While there is an extensive literature on recalibration methods for univariate probabilistic forecasts, work on calibration for multivariate forecasts is much more limited.  
This paper introduces a novel post-hoc recalibration approach that addresses multivariate calibration for potentially misspecified models. Our method involves constructing local mappings between vectors of marginal probability integral transform values and the space of observations, providing a flexible and model free solution applicable to continuous, discrete, and mixed responses. We present two versions of our approach: one uses K-nearest neighbors, and the other uses normalizing flows. Each method has its own strengths in different situations. We demonstrate the effectiveness of our approach on two real data applications: recalibrating a deep neural network's currency exchange rate forecast and improving a regression model for childhood malnutrition in India for which the multivariate response has both discrete and continuous 
components. 
   
\vspace{20pt}
 

\vspace*{\fill}
\noindent {\small\textbf{Acknowledgments:} David Nott's research was supported by the Ministry of Education, Singapore, under the Academic Research Fund Tier 2 (MOE-T2EP20123-0009), and he is affiliated with the Institute of Operations Research and Analytics at the National University of Singapore. Nadja Klein was supported by the Deutsche Forschungsgemeinschaft (DFG, German Research Foundation) through the Emmy Noether grant KL 3037/1-1.}

\vspace{20pt}

\noindent{\small
$^1$ Department of Statistics and Data Science, National University of Singapore, Singapore\\
$^2$ Department of Statistics, University of Brasília, Brazil\\
$^3$ School of Mathematics and Statistics, University of New South Wales, Australia\\
$^4$ Scientific Computing Center, Karlsruhe Institute of Technology, Germany\\
$^\ast$ Correspondence should be directed to guilhermerodrigues@unb.br
}

\end{titlepage}

\spacingset{1.5} 

\section{Introduction}

Historically notions of calibration have their roots in probabilistic forecasting with applications in many fields. 
There are different types of calibration \citep[e.g.,][]{GneBalRaf2007}, but heuristically a model is considered to be calibrated if its predicted uncertainty matches the observed uncertainty in the data in some sense. For example, a forecast might be considered useful for decision making if an event with a certain forecast probability occurs with the corresponding relative frequency.  Calibration of probabilistic models is considered desirable in many scenarios, and has been studied for many types of models and applications, such as neural networks \citep{DheTai2023,LakPriBlu2017}, regression \citep{KleNotSmi2021}, simulator based inference \citep{RodPraSis2018}, and clustering \citep{GuoPleSunKil2017}.
In many situations probabilistic forecasts can be multivariate, involving a vector of random variables $\bm Y=(Y_1,\dots,Y_d)$. However, ensuring multivariate calibration is a challenging task. In this paper, we introduce a novel approach to recalibrating uncertainties obtained from multivariate probabilistic forecasts, based on models which are possibly misspecified.  Our approach can be applied post hoc to arbitrary and 
already fully fitted models ensuring approximate multivariate calibration while simultaneously keeping other properties such as the interpretability of the base model 
intact. While we focus on recalibrating an existing base model, it is possible to use a very flexible model at the outset, and there is an extensive literature on flexible regression beyond the mean \citep{Kne2013,HenZieGne2021}.

To explain our contribution, it is necessary to discuss different types of calibration. For univariate responses, a common choice is probability calibration \citep{GneBalRaf2007}, which can be assessed by checking uniformity of the probability integral transform (PIT) values \citep{Daw1984}. A PIT value is the evaluation of the cumulative distribution function (CDF) of the model at an observed data point. Uniformity of the PIT values implies that prediction intervals derived from the probabilistic model have the correct coverage in a frequentist sense. When extending probability calibration from the univariate to the multivariate setting, it is not enough to check uniformity of PIT values for each marginal separately. 

\citet{Smi1985} considers joint uniformity of the univariate PIT values under a Rosenblatt transformation \citep{Ros1952} summarizing the joint distribution. \citet{DieGunTodTay1998} suggest to check this graphically using histograms and correlograms, and formal tests have been developed in the context of economic forecasting \citep{CorSwa2006,KoPar2013,DovMan2020}. However, this approach is limited, as a Rosenblatt transformation is not readily available for many complex models. 

In the context of ensemble forecasts, \citet{GneStaGriHelJoh2008} introduce multivariate rank histograms as a simple graphical check for multivariate calibration. Multivariate rank histograms are extended to copula PIT (CopPIT) values by \citet{ZieGne2014}. If all univariate marginals of the prediction model are probability calibrated, the CopPIT values depend only on the copula of the forecast. A formal description of this is given in Section~\ref{sec:calibration}. An alternative to PIT and CopPIT values are proper scoring rules \citep{GneRaf2007}, which jointly quantify calibration and sharpness of a probabilistic prediction model. Formal tests for multivariate calibration based on scoring rules complementing PIT-based calibration can be derived \citep{KnuKruPoh2023}.

Even when useful in practice, many models suffer from miscalibration induced by model misspecification. For example, computer based simulators idealise and simplify complex real world phenomena and thus suffer from model misspecification \citep{WarCanBeaFasSch2022}; in regression, highly structured, but therefore misspecified models can be preferable when the focus is on interpretation and not on prediction; modern deep neural networks suffer from several sources of uncertainty that can be challenging to track through their complex structures \citep{GawTasAliLeeHumFenKruTriJunRos2023}. These observations motivate recalibration techniques, which allow to adjust a fitted model post hoc. 

When it comes to recalibration of already estimated models, a rich literature exists in the univariate context. The approaches are often tailored to a specific subclass of statistical models. In the context of parameter estimation, \citet{MenFanGarSis2014} consider recalibration of confidence intervals using bootstrap style samples generated from a predictive distribution under the estimated model. In classification, Platt-scaling \citep{Pla1999}, which extends a trained classifier with a logistic regression model to return class probabilities, is a popular approach. Platt-scaling has been extended in various directions within the machine learning literature \citep[e.g.,][]{GuoPleSunKil2017,KulSilFla2017}. For univariate regression, \citet{KulFenErm2018} suggest learning a transformation of the PIT values to achieve probability calibration.  They use isotonic regression and their approach is extended by \citet{DheTai2023}, who use a kernel density estimator (KDE) of the PIT distribution instead. Our approach can be considered as a multivariate extension to this idea and we review it in more detail in Section~\ref{sec:univariate calibration}.  Our approach is also closely linked to the local recalibration technique for artificial neural networks proposed by \citet{TorNotSisRodReiRod2024}. They use non-parametric PIT transformations on a local neighbourhood learned with K-nearest neighbours (KNN), which can be applied to any layer of a deep neural architecture. 

Despite the clear need, there is still a lack of general multivariate recalibration techniques that consider a vector of quantities of interest jointly in the literature. \citet{HeiHelLenTho2021} discuss post-processing methods for multivariate spatio-temporal forecasting models. However, calibration is only one of their many objectives and the approach does not easily generalize to other model classes. Recently, \citet{WehGamSenBehSapCutJac2024} considered recalibration for simulation-based inference under model misspecification. Their approach involves learning an optimal transport map between real world observations and the output of the misspecified simulator. 

The main contributions of this paper are as follows. (i) We introduce a novel method to achieve multivariate calibration post hoc. The main idea is to construct local mappings between vectors of marginal PIT values and the observation space. Our method thus complements established methods for univariate calibration. (ii) Our approach is general. We are not restricted to continuous data, but can consider discrete and even mixed responses. Therefore the approach can be applied beyond regression to tasks such as clustering, classification, and generalized parameter inference. (iii) Our approach is model-free as we do not assume a particular structure of the underlying base model. Even though it is helpful if the CDFs of the univariate marginals are available in closed form, our method can be applied as long as samples from the base model can be readily generated. (iv) Our method is simple to use. We introduce two versions of our approach. First, a KNN-based approach similar to \citet{TorNotSisRodReiRod2024}, which is then extended to a normalizing flow based approach, where the PIT maps are explicitly learned. Both versions of our approach come with different advantages and we discuss which method is best suited to which scenario. 

We apply our method to two real data examples. First, we recalibrate a one-day ahead forecast for currency exchange rates based on a deep neural network. Multivariate calibration, where all currencies are considered jointly, is desirable due to the complex dependence structure across currencies. Secondly, we consider a regression task concerning childhood malnutrition in India. The bivariate response vector is mixed, containing a continuous and a discrete response. Multivariate recalibration can be used to combine univariate regression models into one joint predictor. Specifying separate regression models for the predictors can be easier then constructing a joint model, especially when working with mixed data. 

The rest of this paper is organized as follows. First, we give some background on different definitions of calibration and existing recalibration techniques in Section~\ref{sec:calibration}. Then, we present our novel recalibration method in Section~\ref{sec:our}. Section~\ref{sec:simulations} illustrates the good performance of our approach for simulated data in a number of scenarios and Section~\ref{sec:applications} considers the aforementioned real data examples. Section~\ref{sec:conclusion} gives a concluding discussion. 

\section{Background on Calibration}\label{sec:calibration}

Let $F(\bm Y,\bm X)$ denote the joint distribution of a response $\bm Y\in \cal Y$ and feature vector $\bm X\in\cal X$. In practice, $F(\bm Y,\bm X)$ is unknown and the conditional distribution $F(\bm Y\mid\bm X)$ is estimated by some probabilistic model $\widehat{F}(\bm Y\mid\bm X)$ from a training set $\mathcal{D}_\text{train}=\{(\bm y_\text{train}^{(i)},\bm{x}_\text{train}^{(i)}),i=1,\dots,n_\text{train}\}$. Heuristically, the model $\widehat{F}$ is said to be calibrated if it correctly specifies the uncertainty in it's own predictions. Since $F$ is not available in practice, calibration can only be assessed based on a validation set $\mathcal{D}_\text{val}=\{(\bm y_\text{val}^{(i)},\bm x_\text{val}^{(i)}),i=1,\dots,n_\text{val}\}$ of observations from $F$, which is potentially disjoint from $\mathcal{D}_\text{train}$. In this section, we will give some background on the notion of calibration by first reviewing univariate calibration in Section~\ref{sec:univariate calibration}, which will be then extended to the multivariate setting in Section~\ref{sec:multivariate calibration}. 

\subsection{Univariate Calibration}\label{sec:univariate calibration}

In the univariate case, where $\cal{Y}\subseteq \mathbb{R}$, several notions of calibration exist within the literature \citep[e.g.,][]{GneRes2023}. Here, we focus on marginal calibration and probability calibration, which are two choices commonly considered in practice.

$\widehat{F}$ is said to be marginally calibrated \citep{GneBalRaf2007} if 
\begin{equation*}
    \mathbb{E}_{\bm{x}\sim F}[\widehat{F}(y\mid \bm{X})]=\mathbb{P}_F(Y\leq y)\qquad \text{for all } y\in\cal Y.
\end{equation*}
That is, the average predictive CDF $\frac{1}{n_\text{val}}\sum_{i=1}^{n_\text{val}}\widehat{F}(y\mid \bm{x}_\text{val}^{(i)})$ matches with the empirical CDF of the observations $\frac{1}{n_\text{val}}\sum_{i=1}^{n_\text{val}}\indicator{y_\text{val}^{(i)}\leq y}$ asymptotically for all $y\in\cal Y$. Hence, \citet{GneBalRaf2007} suggest plotting the average predictive CDF versus the empirical CDF to graphically assess marginal calibration. 

The random variable 
\begin{equation}\label{eq:pit}
    P=\widehat{F}(Y^{-}\mid \bm X)+\Nu\left[\widehat{F}(Y\mid \bm X)-\widehat{F}(Y^{-}\mid \bm X)\right]\qquad\text{for }(Y,\bm X)\sim F(Y,\bm X),
\end{equation}
where $\Nu\sim\UD(0,1)$ and $\widehat{F}(Y^{-}\mid \bm X)$ is the left-handed limit of $\widehat{F}(y\mid \bm X)$ as $y$ approaches $Y$ from below, is the randomized PIT value \citep{CzaGneHel2009}. Note that $P$ depends both on $F(Y,\bm X)$ and the model $\widehat{F}(Y\mid\bm X)$. If $\widehat{F}$ is continuous, $P$ is not randomized $$P=\widehat{F}(Y\mid \bm X).$$ $\widehat{F}$ is said to be probability calibrated if $P\sim\UD(0,1)$. 
For $i=1,\dots,n_\text{val}$ let $\nu^{(i)}$ be independent uniform random variables on $[0,1]$ and write 
\begin{equation}
    p^{(i)} = \widehat{F}(y_\text{val}^{(i)-}\mid \bm x_\text{val}^{(i)})+\nu^{(i)}\left[\widehat{F}(y_\text{val}^{(i)}\mid \bm x_\text{val}^{(i)})-\widehat{F}(y_\text{val}^{(i)-}\mid \bm x_\text{val}^{(i)})\right].
\end{equation}
 $p^{(i)}$ is an empirical evaluation of \eqref{eq:pit} across the validation set. Thus, probability calibration can be graphically checked by plotting a histogram of $\{p^{(i)}, i=1,\dots,n_\text{val}\}$. Formal tests for uniformity of the PIT values based on the Wasserstein distance \citep{ZhoLiWuCar2021,ZhaMaErm2020} and the Cramér-von Mises distance \citep{KulFenErm2018} are popular alternatives to graphical checks. \citet{GneBalRaf2007} show that probability calibration is under mild conditions equivalent to quantile calibration \citep{KulFenErm2018}, which requires 
\begin{equation*}
    \frac{1}{n_\text{val}}\sum_{i=1}^{n_\text{val}}\indicator{y_\text{val}^{(i)}\leq \widehat{F}^{-1}(p\mid\bm x_\text{val}^{(i)})}\to p \qquad \text{almost surely for all } p\in[0,1],
\end{equation*}
 where $\widehat{F}^{-1}(\cdot\mid \bm x_\text{val}^{(i)})$ denotes the generalized inverse of $\widehat{F}(\cdot\mid \bm x_\text{val}^{(i)})$. This perspective has the nice interpretation that prediction intervals derived from $\widehat{F}$ have the correct coverage. 

Histograms of the PIT values can also be used for model criticism as they indicate the type of miscalibration at hand. For example, U-shaped histograms indicate overconfidence, while triangular shapes indicate a biased model \citep{GneBalRaf2007}.
 
Several techniques to recalibrate a potentially miscalibrated model $\widehat{F}$ in a post-hoc step exist in the literature. Here, we describe a simple method for doing this due to \citet{KulFenErm2018}. Write $G(p)$ for the distribution function of the PIT values \eqref{eq:pit}, where dependence on $F$ and $\widehat{F}$ is left implicit in the notation. It is easy to check that $G(\widehat{F}(y\mid\bm x))$ is a distribution function for every $\bm x\in\cal X$ and probability calibrated with respect to $F(Y,\bm X)$. In practice, the distribution function $G(p)$ is not known, and it must be estimated from $\mathcal{D}_\text{val}$. \citet{KulFenErm2018} suggest using a method based on isotonic regression. \citet{DheTai2023} extend this idea and consider KDEs. Among other choices, they propose to use 
\begin{align*}
    \widehat{G}(p)=\frac{1}{n_\text{val}}\sum_{i=1}^{n_\text{val}}\indicator{p_i\leq p},
\end{align*}
which is the empirical CDF from the PIT values over the validation set $\cal{D}_\text{val}$. An alternative approach to recalibration for regression models is given by \citet{SonDieKulFla2019}. Recently, \citet{TorNotSisRodReiRod2024} proposed nonparametric local recalibration for neural networks. Their approach uses a fast KNN algorithm to localize the recalibration and can be used in any layer of the neural network scaling to potentially high-dimensional feature spaces $\cal X$. 

\subsection{Multivariate Calibration}\label{sec:multivariate calibration}
Extending the different notions of calibration from the univariate to the multivariate case, $\mathcal{Y}\subseteq\mathbb{R}^d$, is not straightforward. One reason for this is that the multivariate integral transformation
\begin{equation*}
    F(\bm y\mid\bm x)\qquad\text{for}\qquad(\bm y, \bm x)\sim F(\bm Y,\bm X)
\end{equation*}
is, in contrast to the univariate case, generally not uniformly distributed \citep[e.g.,][]{GenRiv2001}, but follows the so-called Kendall distribution of $F$. The Kendall distribution depends only on the copula of the multivariate probability measure, and thus summarizes the dependence structure of $F$. Based on this observation, \citet{ZieGne2014} introduce copula probability integral transform (CopPIT) values as analogous to the univariate PIT values described in \eqref{eq:pit}. The CopPIT values are given as
\begin{equation}\label{eq:coppit}
    U = \mathcal{K}_{\bm X}\left(\widehat{F}(\bm Y^{-}\mid \bm X)\right)+\Upsilon\left[\mathcal{K}_{\bm X}\left(\widehat{F}(\bm Y\mid \bm X)\right)-\mathcal{K}_{\bm X}\left(\widehat{F}(\bm Y^{-}\mid \bm X)\right)\right], 
\end{equation}
where $\Upsilon\sim\UD(0,1)$, $(\bm Y,\bm X)\sim F(\bm Y,\bm X)$, and $\mathcal{K}_{\bm X}$ denotes the Kendall distribution of $\widehat{F}(\bm Y\mid \bm X)$. $\widehat{F}$ is said to be copula calibrated if the CopPIT values are uniformly distributed on the unit interval \citep{ZieGne2014}. In this way, copula calibration can be seen as a multivariate extension to probability calibration. In particular for $d=1$, $\mathcal{K}_{\bm X}$ is the uniform distribution on $[0,1]$, so that \eqref{eq:coppit} is equal to $\eqref{eq:pit}$. As in the univariate case, let 
\begin{equation*}
    u^{(i)} = \mathcal{K}_{\bm x_\text{val}^{(i)}}\left(\widehat{F}(\bm y_\text{val}^{(i)-}\mid \bm x_\text{val}^{(i)})\right)+\upsilon^{(i)}\left[\mathcal{K}_{\bm x_\text{val}^{(i)}}\left(\widehat{F}(\bm y_\text{val}^{(i)}\mid \bm x_\text{val}^{(i)})\right)-\mathcal{K}_{\bm x_\text{val}^{(i)}}\left(\widehat{F}(\bm y_\text{val}^{(i)-}\mid \bm x_\text{val}^{(i)})\right)\right], 
\end{equation*}
denote the empirical CopPIT values from the validation set, $i=1,\dots,n_\text{val}$, where $\upsilon^{(i)}$ are independent uniform variates on $[0,1]$. Again, copula calibration can be assessed by checking uniformity of $\{u^{(i)}, i=1,\dots,n_\text{val}\}$. However, interpretation of the CopPIT histograms is more challenging than in the univariate case, as they not only summarize potential miscalibration of the dependence structure, but also of the marginal distributions. However, in the special case that all margins of $\widehat{F}$ are uniformly probability calibrated, the CopPIT values summarize miscalibration of the copula of $\widehat{F}$ only \citep{ZieGne2014}. Thus, in practice it is sensible to assess multivariate calibration by checking for univariate calibration of each marginal in terms of the marginal PIT values \eqref{eq:pit} and copula calibration in terms of the CopPIT values \eqref{eq:coppit}.  

\citet{ZieGne2014} also introduce Kendall calibration 
\begin{equation}\label{eq:kendall_calibration}
    \lim_{n_\text{val}\to\infty} \frac{1}{n_\text{val}}\sum_{i=1}^{n_\text{val}}\indicator{ \widehat{F}(\bm y^{(i)}_\text{val}\mid \bm x^{(i)}_\text{val}) \leq\omega} =  \lim_{n_\text{val}\to\infty}  \frac{1}{n_\text{val}}\sum_{i=1}^{n_\text{val}} \mathcal{K}_{\bm x_\text{val}^{(i)}}(\omega)\qquad \text{for all } \omega\in[0,1]
\end{equation}
as the multivariate analogue to marginal calibration. Kendall calibration can be assessed by a so called Kendall diagram, which is a scatter plot of the empirical left hand side versus the empirical right hand side of \eqref{eq:kendall_calibration} for different values of $\omega$. 

Both copula calibration and Kendall calibration necessitate the derivation of the Kendall distributions $ \mathcal{K}_{\bm x}$. The Kendall distribution can be calculated in closed form only for a few special cases \citep[e.g.,][]{GenRiv2001}. So, in practice, $\mathcal{K}_{\bm x}$ in \eqref{eq:coppit} and \eqref{eq:kendall_calibration} is replaced by an approximation given as the empirical CDF of the pseudo observations \citep{BarGenGhoRem1996}
\begin{equation*}
    w_k=\frac{1}{m}\sum_{j=1}^m \indicator{ \bm y_j\preceq \bm y_k} \qquad\text{for }k=1,\dots,n,
\end{equation*}
where $\bm y_j=(y_{j1},\dots,y_{jd})\preceq \bm y_k=(y_{k1},\dots,y_{kd})$ if $y_{jl}\leq y_{kl}$ for all $l=1,\dots,d$ and $\bm y_1,\dots,\bm y_m$ is a large sample from $\widehat{F}(\bm y\mid\bm x)$. 

\section{Multivariate Calibration via PIT mapping}\label{sec:our}

This section describes our approach to recalibrate arbitrary probabilistic prediction models $\widehat{F}$. We consider a simple KNN approach, which can be thought of as a multivariate extension to the recalibration methods by \citet{TorNotSisRodReiRod2024} and \citet{RodPraSis2018} first in Section~\ref{sec:ourKNN}, and then the novel normalizing flow based method in Section~\ref{sec:ourNF}.

\subsection{Nearest neighbour recalibration}\label{sec:ourKNN}

Suppose that we have a mapping on the feature space, $h:\mathcal{X}\to\mathbb{R}^d$. The purpose of the function $h$ is to reduce the dimension of $\bm x$ and we use $\norm{h(\bm{x})-h(\bm{x}')}$ to measure the similarity of the feature vectors $\bm x$ and $\bm x'$. Let
\begin{equation*}
    N_k(\bm x)=\{i: h(\bm{x}^{(i)}_\text{val})\text{ is one of the $k$ nearest neighbours of }h(\bm x)\}.
\end{equation*}

If $N_k(\bm x)$ is a sufficiently small neighbourhood around $\bm x$, $\{\bm p^{(i)}, i\in N_k(\bm x)\}$ approximates a sample from $\bm P=(P_1,\dots,P_d)$, where $P_l$ is the PIT value for the $l$-th response of $\widehat{F}(\bm Y\mid \bm x)$ as given in \eqref{eq:pit}. Let $G_{\bm x}$ denote the joint distribution of $\bm P$ given $\bm x$ with marginal distributions $G_{\bm x,l}$, $l=1,\dots,d$. Theoretical properties of $G_{\bm x}$ were studied in \citet{RodPraSis2018}. In particular, $G_{\bm x}(\widehat{F}(\bm y\mid\bm x))=\left(G_{\bm x,1}(\widehat{F}_1(y_1\mid\bm x)),\dots,G_{\bm x,d}(\widehat{F}_d(y_d\mid\bm x))\right)$ has probability calibrated marginals following the same arguments as for the univariate recalibration techniques described in Section~\ref{sec:univariate calibration}. Note that the use of $G_{\bm x,l}$ instead of the global unconditional distribution $G_l$ as considered in \citet{KulFenErm2018} and \citet{DheTai2023} gives a stronger form of calibration, as the resulting model is locally, that is conditional on $\bm x$, probability calibrated. In addition to the marginal information, $G_{\bm x}$ also matches the dependence structure of $\widehat{F}(\bm Y\mid\bm x)$ under $F(\bm Y\mid\bm x)$. For a given $\bm x \in \cal X$ and continuous marginals $\widehat{F}_l(y\mid\bm x)$, $p_l$ is a non-random transformation of $y_l$ and, in particular, $\bm p$ is an invertible transformation of $\bm y$. The Kendall distribution is invariant under such transformations and thus the CopPIT value $u\mid \bm x$ could be calculated purely on $\bm p\mid\bm x$, without access to $\bm y\mid\bm x$. Also, $\bm P\mid \bm x$ has copula $C_{\bm x}$, which is the copula of $F(\bm Y\mid \bm x)$ and, from the arguments above, $G_{\bm x}(\widehat{F}(\bm Y\mid\bm x))$ has probability calibrated marginals. Following \citet{ZieGne2014} this implies copula calibration.

Thus, for a given $\bm x\in\cal X$, a sample of size $k$ from an approximately calibrated predictive distribution $\widetilde{F}(\bm Y\mid \bm x)$ can be generated as 
\begin{equation} \label{eq:knn_approximation}
    \tilde{\bm y}^{(i)} = \left(\tilde{y}^{(i)}_1,\dots,\tilde{y}^{(i)}_d\right) = \left(\widehat{F}^{-1}_1(p^{(i)}_1\mid\bm x),\dots,\widehat{F}^{-1}_d(p^{(i)}_d\mid\bm x)\right) = \widehat{F}^{-1}(\bm p^{(i)}\mid\bm x),\quad i\in N_k(\bm x).
\end{equation}
Here, $\bm p^{(i)}=(p_1^{(i)},\dots,p_d^{(i)})$ with $p^{(i)}_l$ the empirical PIT value for the $l$-th marginal distribution $\widehat{F}_l(Y_l\mid \bm X)$ evaluated on the $i$-th entry of the validation set. If $\widehat{F}^{-1}_l(\cdot\mid\bm x)$ is not available in closed form it can be easily approximated using a sample from $\widehat{F}_l(\bm Y\mid\bm x)$ making our approach model free.

However, \eqref{eq:knn_approximation} is only an approximation to $G_{\bm x}(\widehat{F}(\bm y\mid\bm x))$ and we will illustrate how well this works in practice on a number of simulated and real data examples in Sections~\ref{sec:simulations} and~\ref{sec:applications}.

\subsection{Recalibration with normalizing flows}\label{sec:ourNF}
We can think of the nearest neighbour approach introduced in Section~\ref{sec:ourKNN} as obtaining an approximate sample from $\bm P\mid \bm X=\bm x$ for a target feature vector $\bm x$, and then transforming back to the original space of the responses to obtain approximate samples of $\bm Y\mid\bm X=\bm x$. In the nearest neighbour approach, no explicit expression for $\widetilde{F}(\bm Y\mid \bm x)$ is constructed, and the number of potential draws from $\widetilde{F}(\bm Y\mid \bm x)$ is restricted by $k$ heavily depending on $n_\text{val}$. However, some applications require calculating complex summary statistics from the potentially intricate distribution $\widetilde{F}$, which necessitates the ability to draw arbitrary large samples from the recalibrated model. Thus, we propose a similar method to the KNN approach using normalizing flows to draw approximate samples from $\bm P\mid \bm x$. 

The basic idea is as follows. Let $\rho(\bm z)$ be a reference density with respect to the Lebesgue measure on $\mathbb{R}^d$, which we take to be the standard normal density. We consider a bijective transformation $T_{\bm\zeta}(\bm z\mid\bm x)$, and transform $\bm Z\sim\rho(\bm z)$ to a random vector $\bm P\mid\bm x$, where $\bm\zeta$ is a set of learnable parameters. 
The density of $\bm P\mid\bm x$ is thus approximated as 
\begin{equation*}
    \rho\left(T^{-1}_{\bm\zeta}(\bm p\mid\bm x)\right)\vert\det J_{T^{-1}_{\bm\zeta}}(\bm p\mid\bm x)\vert,
\end{equation*}
where $T^{-1}_{\bm\zeta}(\bm p\mid\bm x)$ is the inverse of $T_{\bm\zeta}(\bm z\mid\bm x)$, and $J_{T^{-1}_{\bm\zeta}}(\bm p\mid\bm x)$ is its Jacobian matrix. Based on this, the parameter ${\bm\zeta}$ is learned using observations $(\bm p^{(i)},\bm x^{(i)}_\text{val})$, where $\bm p^{(i)}$ denotes the vector of PIT values for the marginal distributions evaluated on the validation set. To avoid boundary effects, we consider normalized PIT values $\bm p_N=\Phi^{-1}(\bm p)=\left(\Phi^{-1}(p_1),\dots,\Phi^{-1}(p_d)\right)\in\mathbb{R}^d$, where $\Phi(\cdot)$ is the CDF of the standard Gaussian distribution, instead of the usual PIT values on $[0,1]$. As for the KNN approach, $\bm x$ can be replaced with a lower dimensional representation $h(\bm x)$ in the construction of $T_{\bm\zeta}(\bm z\mid\bm x)$. Having learned ${\bm\zeta}$ as $\hat{{\bm\zeta}}$, samples from $\bm P\mid\bm x$ can be generated by sampling $\bm z\sim\rho(\bm z)$ and setting $\bm p = T_{\hat{{\bm\zeta}}}(\bm z\mid\bm x)$. These samples can be then used similar to \eqref{eq:knn_approximation} to generate samples from the approximately recalibrated predictive distribution $\widetilde{F}(\bm Y\mid\bm x)$. Pseudo code for the full approach is given in Appendix A.

There are many ways to construct suitable transformations $T_{{\bm\zeta}}(\bm z\mid\bm x)$ in the literature on transport maps \citep[e.g.,][]{MarMosParSpa2016} and normalizing flows \citep[e.g.,][]{RipAda2013,YaoLiyLoTseChaLee2023} as well as standard software for conditional density estimation. Here, we consider the real-valued non-volume preserving (real-NVP) approach by \citet{DinSohBen2016}, where $T_{{\bm\zeta}}(\bm z\mid\bm x)$ is given as a stack of affine coupling layers. We found this to be a satisfactory choice in all examples considered, but alternative approaches might be better suited depending on the structure of the recalibration task at hand.

The normalizing flow can be interpreted as a conditional density estimate for $\bm P\mid\bm x$. If $d=1$, our approach can therefore be considered a localized version of the KDE based method by \citet{DheTai2023}. 

\section{Simulations} \label{sec:simulations}

We illustrate the performance of both the KNN based approach labeled KNN, and the normalizing flow approach labeled NF on a number of simulated examples.
First, we reanalyze the illustrative example from \citet{ZieGne2014} to consider forecasts suffering different kinds of miscalibration. Both KNN and NF achieve probability calibration of the marginals, copula calibration and Kendall calibration across all scenarios.
Secondly, we consider a regression task, where $\bm Y\mid \bm X$ is degenerate and we investigate the local calibration properties of our approaches. For a given $\bm x$, both NF and KNN allow to generate samples from the recalibrated predictive distribution $\bm Y\mid \bm x$. KNN can generate only a small sample of size much smaller then $n_\text{val}$, while NF is computational more complex, but allows to draw an arbitrarily large sample from the recalibrated model. In our simulations, samples from both methods are close to the true distribution even for a grossly misspecified base model. More details on the simulation studies can be found in Appendix~B. 

\section{Applications}\label{sec:applications}
\subsection{Currency exchange rates}

Foreign currencies constitute a popular class of assets among investors. In so-called Forex trading, traders exchange currencies with the goal of making a profit. The ability to make reliable predictions for currency exchange rates is crucial for a successful trading strategy. To this end, we analyze five time series of daily exchange rates for five currencies relative to the US dollar: the Australian Dollar (AUD), the Chinese Yuan (CNY), the Euro (EUR), the Pound Sterling (GBP), and the Singapore Dollar (SGD). These data span five years from August 01, 2019, to August 01, 2024, and were sourced from Yahoo Finance. Given the high correlation among currency exchange rates, multivariate calibration is a desirable feature for any currency exchange forecasting model.

\paragraph{Baseline model} We consider a one-day ahead forecast based on a Long Short-Term Memory (LSTM) Neural Network with a distributional layer, so that the resulting forecast distribution is multivariate Gaussian with diagonal covariance structure. Even though the resulting probabilistic forecast cannot express correlation, dependencies between the currencies are exploited through the deep LSTM network modelling the $5$-dimensional time-series jointly. LSTMs have been successfully implemented for time-series prediction \citep[e.g.,][]{HuaZhaLiCheLiuZha2019} and the resulting model recovers the general structure of the data well. Note however that our main focus is to illustrate the merits of multivariate calibration and not on the construction of the forecasting model.

\paragraph{Recalibration in online learning} 
Every day, as a new data point becomes available, the baseline LSTM model is updated accordingly. The recalibration model follows a similar iterative process. After an initial period (here 100 days), the process is as follows. The LSTM model generates a predictive distribution for the one-day-ahead forecast. This forecast is then recalibrated using the recalibration model. Once the actual data for the next day is obtained, the LSTM model is updated on the now extended dataset. Simultaneously, the new data point yields an updated vector of marginal PIT values, prompting an update to the recalibration model. In each step only one additional data point becomes available, and both the base and the recalibration model can be updated using a warm-start avoiding the need to retrain the models from scratch. This drastically reduces the computational resources needed for training. We do not use a separate validation set, but reuse the training data for recalibration of the forecast. 

In time series forecasting, a natural assumption is that the more recent an observation was made, the more information it contains on future values. Hence, here we consider the KNN approach, where we use the most recent 100 PIT values for recalibration at each time step. This way, the recalibration is carried out on a rolling window of PIT values. The KNN approach is preferable to NF here as it allows us to gradually control the information available to the recalibration method. 

\begin{figure}[bt!]
    \centering
    \includegraphics[width=\textwidth,keepaspectratio]{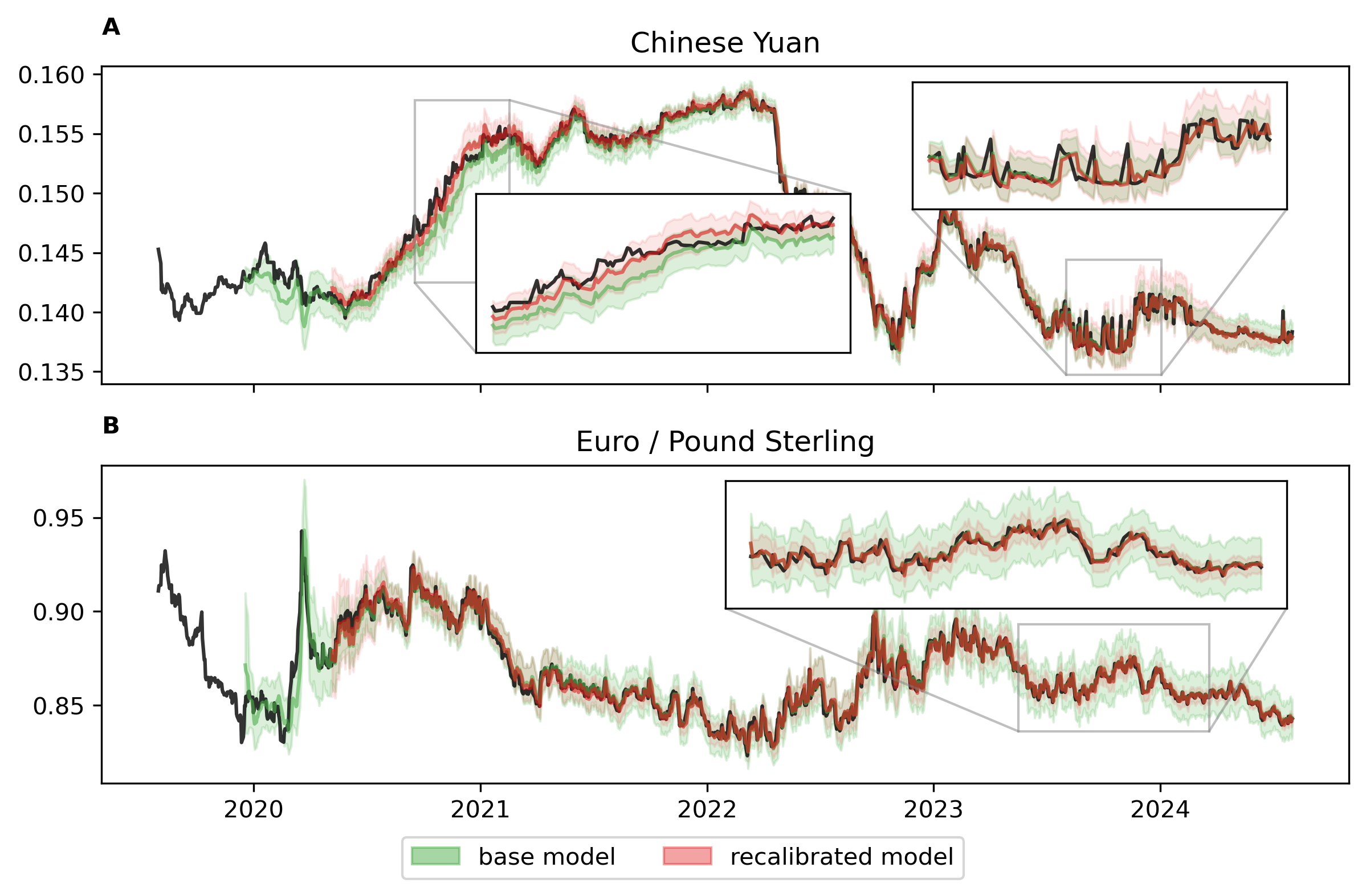}
    \caption{\small Currency exchange. One day ahead forecasts for CNY \textbf{(A)} and EUR/GBP \textbf{(B)} under the base model (green) and the recalibrated model (red). The bold lines correspond to the estimated mean values while $95\%$ credible bands are given by the shaded area. The true realized time series are given in black.}
    \label{fig:cur2}
\end{figure}

\paragraph{Results} Histograms of the univariate PIT values (Appendix~C) indicate that none of the margins of the base model are probability calibrated, and the kind of miscalibration differs drastically between currencies. For example, the marginal PIT values for CNY and SGD are skewed indicating a biased forecast, while the histograms for AUD and EUR are U shaped indicating underdispersion \citep{GneBalRaf2007}. The recalibration through KNN drastically improves the overall calibration of the base model. Figure~\ref{fig:cur2}A shows the base, and the recalibrated forecast together with the realized values for CNY. The base model underestimates the exchange rate for CNY in 2020 and 2021. This bias is corrected by the recalibration. In the third and fourth quarter of 2023, the base model underestimates the uncertainty of the forecast, resulting in an accumulation of realised values outside of the $95\%$ credible band during this time period. Our KNN approach detects this local miscalibration and widens the credible band in this time period. 
Multivariate calibration is especially helpful when estimating functions over multiple margins, as done for example when assessing the risk of a portfolio. To illustrate this, we consider the time-series EUR/GBP, which gives the direct exchange rate for EUR relative to GBP. Both the base and the recalibrated forecast model describe an implicit forecast for EUR/GBP. EUR and GBP are strongly correlated with an estimated Kendall's $\tau$ of $0.68$.  Since the base model does not account for this dependence structure, the estimated credible intervals are wider than necessary. Even though KNN does not recalibrate EUR/GBP directly, this is corrected by the multivariate recalibration as shown in Figure~\ref{fig:cur2}B.

\subsection{Childhood malnutrition}

Ending all forms of malnutrition is one of the sustainable development goals of the United Nations \citep{UN2015}. Here, we consider a sample from the Demographic and Health surveys (\url{www.measuredhs.com}) containing $n=24,286$ observations on several factors of undernutrition in India. Following previous analyses of the data \citep{KleKneMarRad2020,SanKleKliMay2024} we consider two responses. The continuous indicator \texttt{wasting} reports weight for height as a z-score and the binary response \texttt{fever} indicates fever within the two weeks prior to the interview. Following \citet{KleKneMarRad2020} we consider the covariables \texttt{csex} indicating the sex of the child, \texttt{cage} the age of the child in months, \texttt{breastfeeding} the duration of breastfeeding in months, \texttt{mbmi} the body mass index of the mother, and \texttt{dist} the district in India the child lives in. 

\paragraph{Baseline model} We fit separate regression models for the two responses. \texttt{wasting} is modelled through a heteroscedastic Gaussian distribution, where both the mean and the variance parameter are linked to an additive predictor, and we use logistic regression for \texttt{fever}. Non-linear effects for the continuous covariates \texttt{cage}, \texttt{breastfeeding}, and \texttt{mbmi} are modelled with Bayesian P-splines \citep{EilMar1996}. We use a linear effect for \texttt{csex}, and a spatial effect with a Gaussian Markov random field prior for \texttt{dist} \citep{RueHel2005}. This results in two highly interpretable distributional regression models. 

\begin{figure}[bt!]
    \centering
    \includegraphics[width=\textwidth,keepaspectratio]{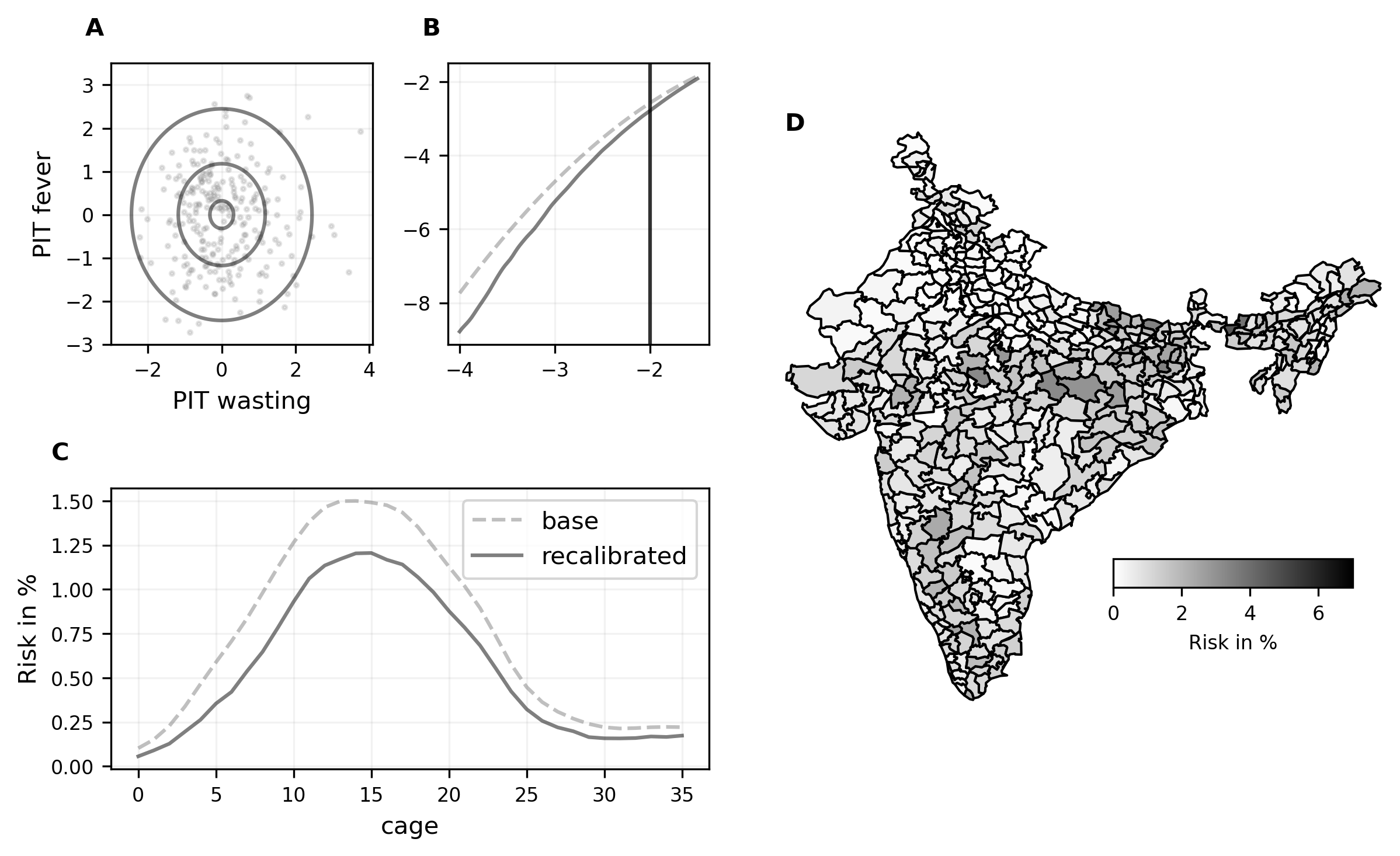}
    \caption{\small Malnutrition. \textbf{A)} Scatter plot of the normalized PIT values with respect to the base model. A contour plot of the bivariate standard Gaussian distribution, which is the reference distribution for NF, is given in grey. \textbf{B)} Log-density for \texttt{wasting} for the base model (dashed) and the recalibrated model (bold). The cut-off value for wasting according to the WHO definition is indicated by the vertical line. \textbf{C)} Main effect for \texttt{cage} for the base model (dashed) and the recalibrated model (bold). The y-axis denotes the risk of a child suffering from wasting and fever in \%. \textbf{D)} Main effect for \texttt{dist} under the recalibrated model. The risk of a child suffering simultaneously from wasting and fever in \% is indicated by the shade of the region. A darker shade corresponds to a higher risk.}
    \label{fig:undernutrition}
\end{figure}

\paragraph{Multivariate calibration} Figure~\ref{fig:undernutrition}A shows a scatter plot of the normalized PIT values $\bm p^{(i)}_N$ for the independent baseline regression models. While the PIT values for \texttt{fever} under the logistic regression model are close to the uniform distribution, the PIT values for \texttt{wasting} show clear deviations from uniformity especially in the upper tail. Since both \texttt{fever} and \texttt{wasting} are indicators of the child's health, a complex dependence structure between the two response variables that is not sufficiently accounted for by the baseline model is expected. Multivariate calibration is used to combine the two univariate distributional regression models into a single multivariate regression model. Since we want to investigate how the recalibration affects the interpretable effects of the univariate regression models, we need to be able to generate large samples from $p(\bm y\mid \bm x)$, which is not possible with the basic KNN approach. We will thus consider the NF approach as described in Section~\ref{sec:our}. We condition the NF on the continuous covariables \texttt{cage}, and \texttt{mbmi} as they are predominant factors in the marginal regression models. 

\paragraph{Results} The NF improves both the probabilistic calibration of the marginals and the copula calibration of the bivariate regression model as described by the PIT and CopPIT values respectively (Appendix~C). The World Health Organization defines a child suffering from wasting if $\texttt{wasting}\leq-2$. Figure~\ref{fig:undernutrition}B shows the left tail for the predictive density of \texttt{wasting} conditional on the median values for all covariables for both the baseline and the recalibrated model. The baseline model overestimates the risk of the median child suffering wasting $\text{Pr}(\texttt{wasting}\leq-2\mid \bm x)$ compared to the recalibrated model. 

The joint regression model allows us to study the risk of a child having fever and simultaneously suffering from wasting $\text{Pr}(\texttt{wasting}\leq-2,\texttt{fever}=1)$. Figures~\ref{fig:undernutrition}C and D show the main effects of \texttt{cage} and \texttt{dist} on this risk respectively. The main effects are calculated by varying the covariable of interest, while keeping the other covariables fixed to their median values. The risk increases for children younger than a year and decreases for older children. The likelihood for \texttt{fever} is increasing for $0\leq\texttt{cage}\leq 12$ according to the baseline logistic regression model. Both the baseline and the recalibrated model find similar shapes for the main effect for \texttt{cage}, but in terms of magnitude the recalibrated model estimates a lower risk. Similarly, the estimated main effect for \texttt{dist} is lower for the recalibrated model than for the baseline model in all 438 districts (Appendix~C). According to this analysis the risk of a child suffering simultaneously from wasting and fever is higher in the mid-eastern districts of India. This is consistent with the findings in \citet{SanKleKliMay2024}.
Even though the recalibrated model is nonparametric, the interpretability of the baseline regression models can be maintained, making the multivariate recalibration approach a valuable tool for the development of complex multivariate distributional regression models. 

\section{Conclusion and Discussion} \label{sec:conclusion}
In this paper, we have introduced a novel approach for recalibrating multivariate models, addressing a critical gap in the calibration literature. Our method involves local mappings between marginal PIT values and the space of the observations and extends established univariate recalibration techniques to the multivariate case. We discuss two different versions of our approach. The KNN-based method provides simplicity and ease of implementation, but is limited as it only allows the generation of a small sample from the recalibrated model. While being computationally more challenging, the NF-based method overcomes this limitation. The merits of our approach are illustrated on a number of simulated and real data examples. We consider forecasting of a multivariate time series and regression for mixed data, further illustrating the versatility of our approach.
However, theoretical properties of the PIT-based mappings are not well investigated. A better theoretical understanding could potentially lead to improved recalibration techniques. We use transformations of the PIT values from the marginal distributions. However, depending on the structure of the underlying predictor model, other univariate distributions that summarize the joint distribution could be considered. 
Future research could investigate the application of our recalibration technique to additional model types, including more intricate dependence structures and larger datasets. Additionally, our method focuses purely on calibration, and integrating it with other aspects of model evaluation and improvement, such as sharpness and robustness checks, could enhance its utility. 
Calibration is also an important tool for model criticism. The local nature of our approach could potentially allow to detect areas of model misspecification and thus our approach could be developed further into model specification and model selection pipelines. 
\FloatBarrier
\bibliography{bib}
\FloatBarrier
\section*{Appendix}
\renewcommand{\thefigure}{\thesection.\arabic{figure}}
\appendix
\section{Pseudo Code}

Pseudo code for the NF approach is given in Algorithm~\ref{alg:nf}.

\RestyleAlgo{ruled}
\begin{algorithm}[bt!]
\caption{Recalibration with normalizing flows}\label{alg:nf}
\textbf{A: Calculate normalized PIT values}\\
\KwIn{ Validation set $\mathcal{D}_\text{val}=\{(\bm y_\text{val}^{(i)},\bm x_\text{val}^{(i)}),i=1,\dots,n_\text{val}\}$; CDF-valued predictive distribution $\widehat{F}(\bm Y\mid\bm X)$ with marginals $\widehat{F}_1(Y_1\mid\bm X),\cdots,\widehat{F}_D(Y_1\mid\bm X)$;
}
\For{$i\gets1$ \KwTo $n_\text{val}$}{
\For{$l\gets1$ \KwTo $d$}{
Sample $\nu^{(i)}_l\sim\UD(0,1)$\;
Set $p^{(i)}_l = \widehat{F}_l\left(y_{\text{val},l}^{(i)-}\mid \bm x_\text{val}^{(i)}\right)+\nu_l^{(i)}\left[\widehat{F}_l\left(y_{\text{val},l}^{(i)}\mid \bm x_\text{val}^{(i)}\right)-\widehat{F}\left(y_{\text{val},l}^{(i)-}\mid \bm x_\text{val}^{(i)}\right)\right]$
}
Let $\bm p_N^{(i)}=\left(\Phi^{-1}\left(p^{(i)}_1\right),\dots,\Phi^{-1}\left(p^{(i)}_d\right)\right)$ be the vector of normalized PIT values
}
\BlankLine
\textbf{B: Train normalizing flow}\\
\KwIn{ Data $\left\{(\bm p_N^{(i)},\bm x^{(i)}_\text{val}), i=1,\dots,n_\text{val}\right\}$; an invertible map $T_{\bm\zeta}(\bm z\mid\bm x)$;
}
Set $\hat{{\bm\zeta}}=\arg\max_{\bm\zeta} \prod_{i=1}^{n_\text{val}}\rho\left(T^{-1}_{\bm\zeta}(\bm p^{(i)}\mid\bm x_\text{val}^{(i)})\right)\vert\det J_{T^{-1}_{\bm\zeta}}(\bm p^{(i)}\mid\bm x_\text{val}^{(i)})\vert$\;
\BlankLine
\textbf{C: Sample from the recalibrated model}\\
\KwIn{ An invertible map $T_{\bm\zeta}(\bm z\mid\bm x)$ with trained parameter $\hat{{\bm\zeta}}$; an observation $\bm x_\text{obs}$; number of samples to be drawn $n$;
}
\For{$j\gets1$ \KwTo $n$}{
Sample $\bm z^{(j)}=\left(z^{(j)}_1,\dots,z^{(j)}_d\right)\sim\ND_d(0,I_d)$\;
Set $\tilde{\bm p}_N^{(j)}=\left(\tilde{p}^{(j)}_{N,1},\dots,\tilde{p}^{(j)}_{N,d}\right)=T_{\hat{{\bm\zeta}}}(\bm z^{(j)}\mid\bm x_\text{obs})$\;
\For{$l\gets1$ \KwTo $d$}{
Set $\tilde{y}^{(j)}_l=\widehat{F}^{-1}_l\left(\Phi\left(\tilde{p}^{(j)}_{N,l}\right)\mid\bm x_\text{obs}\right)$\;
}
Set $\tilde{\bm y}^{(j)}=\left(\tilde{y}^{(j)}_1,\dots,\tilde{y}^{(j)}_d\right)$\;
}
Return $\left\{\tilde{\bm y}^{(j)}, j=1,\dots,n\right\}$\;
\end{algorithm}

\section{Simulations}\label{app:simulations}

This section contains detailed results for the simulation studies.

\subsection{Bivariate copula model}
To illustrate how our proposed approach handles different miscalibrated forecasts, we reanalyze the illustrative example from \citet{ZieGne2014}. The true data generating process (DGP) is a bivariate distribution with normal margins and a Gumbel copula with parameters $\bm \theta=\left(\mu_1,\sigma_1^2,\mu_2,\sigma_2^2,\tau\right)$, where $\mu_j$ is the mean and $\sigma_j^2$ the variance for the $j$-th marginal, $j=1,2$, and $\tau$ is Kendall's $\tau$ parameterizing the Gumbel copula. $\sigma_1^2=1$, $\mu_2=0$ are fixed and the remaining parameters depend on a bivariate vector of covariates $\bm x=(x_1,x_2)$ following independent beta distributions $x_1\sim\BetaD(2,5)$, $x_2\sim\BetaD(5,2)$. Under the true DGP, $\mu_1=2-x_1$, $\sigma_2^2=x_2^{-1}$ and $\tau=\frac{x_1+x_2}{2}$. All forecasts considered specify a Gumbel copula with Gaussian marginals, but potentially misspecify the parameter vector $\bm\theta\equiv\bm\theta(\bm x)$. We consider all $8$ possible combinations of the following three fallacies.
\begin{itemize}
    \item The forecast distribution for the first marginal is either correctly specified (T) or biased $\mu=0.8(2-x_1)$ (F).
    \item The forecast distribution for the second marginal is either correctly specified (T) or underdispersed $\sigma_2^2=0.8x_2^{-1}$ (F).
    \item Kendall's $\tau$ is either correctly specified (T) or underestimated $\tau=0.6\frac{x_1+x_2}{2}$ (F).
\end{itemize}
As in \citet{ZieGne2014}, we denote each of the forecasts by a combination of three letters, where the first letter denotes if the first margin is misspecified, the second letter denotes if the second margin is misspecified and the last letter denotes if the copula is misspecified. For example, FFT denotes the forecast with misspecified margins, but correctly specified dependence structure. 

We consider a validation set with $n=4,000$ samples and evaluate the performance on a hold-out test set of $4,000$ samples. Due to the fixed structure of the forecasting models there is no training set used here. We compare both the KNN approach and the NF approach. For KNN we use the $5\%$ nearest samples according to the Euclidean norm in covariate space. 

\begin{figure}
    \centering
    \includegraphics[width=\textwidth,keepaspectratio]{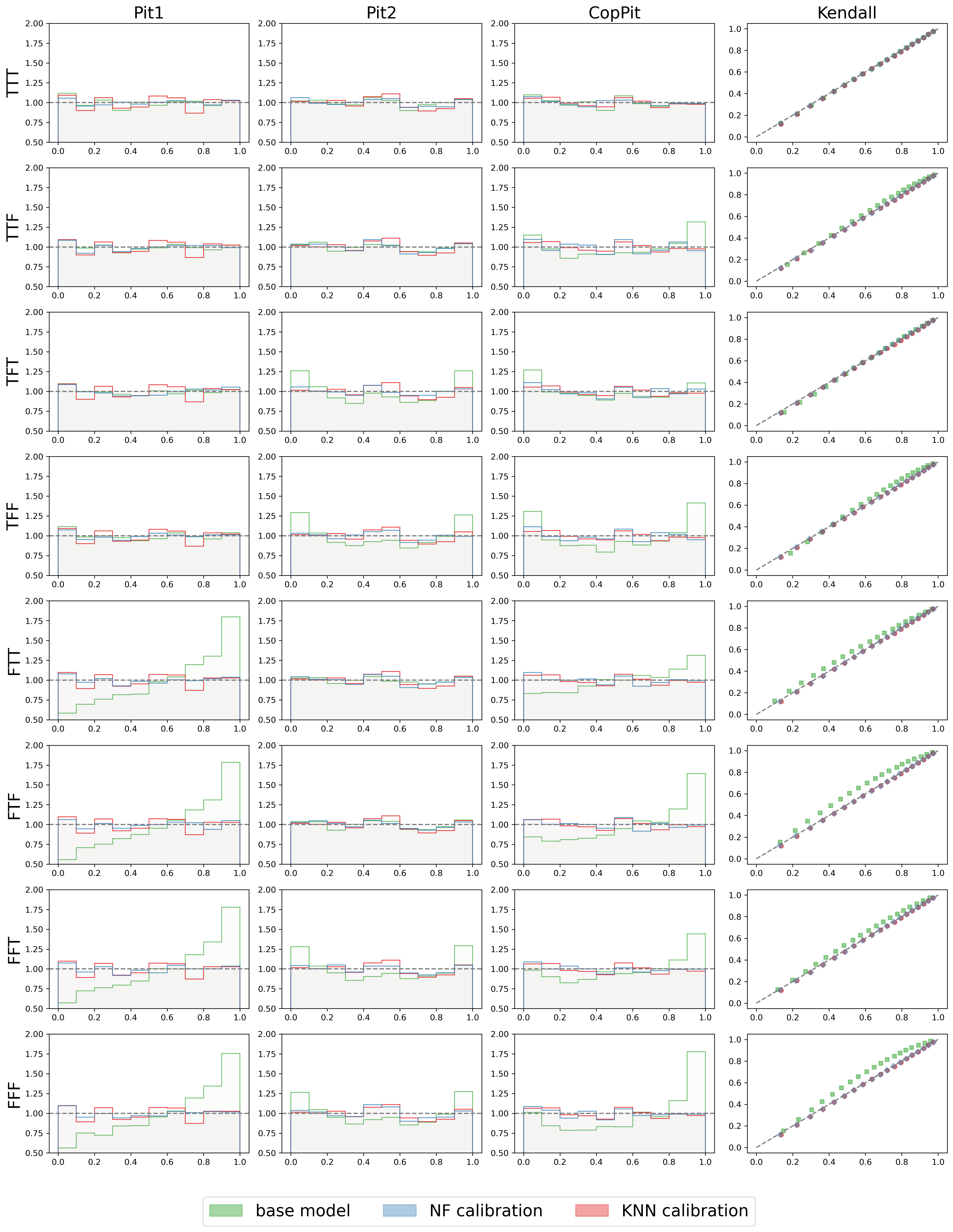}
    \caption{\small Simulations. Performance across the $8$ different forecast specifications (rows) for the uncalibrated base model (green), NF recalibration (blue), and KNN recalibration (red). The first two columns show histograms for the marginal pit values, the middle column shows histograms for the CopPit values, and the fourth column is the Kendall plot.}
    \label{fig:sim}
\end{figure}

Figure~\ref{fig:sim} summarises the results. KNN and NF perform very similar. Both approaches achieve probability calibration of the marginals as summarized through histograms of the univariate PIT values (Columns 1+2 of Figure~\ref{fig:sim}), copula calibration (Column 3 of Figure~\ref{fig:sim}) and Kendall calibration as indicated by the Kendall plot (Column 4 of Figure~\ref{fig:sim}). The forecast TTT is optimal in the sense that the forecast matches the true DGP exactly and neither NF nor KNN seem to deteriorate the forecast.

\subsection{Twisted Gaussians}
We consider the following DGP inspired by a related example in \cite{RodPraSis2018}:  
\begin{align*}
    X&\sim\ND(0,1)\\
    Y_1\mid X&\sim\ND(0,1)\\
    Y_2\mid Y_1,X &\sim Y_1+XY_1^2,
\end{align*}
with $\bm Y=(Y_1,Y_2)$, from which we draw $n_\text{train}=5,000$ samples to train the base model, and $n_\text{val}=5,000$ samples as the validation set. The base model consists of two univariate, Gaussian, linear, homoscedastic models $y_j\mid x \sim \ND(\beta_{0j}+x\beta_{1j},\sigma_j^2)$, $j=1,2$. Hence, the marginal $Y_1\mid X$ of the base model is approximately probability calibrated, while the second marginal and the dependence structure are miscalibrated. 

\begin{figure}[bt!]
    \centering
    \includegraphics[width=\textwidth,keepaspectratio]{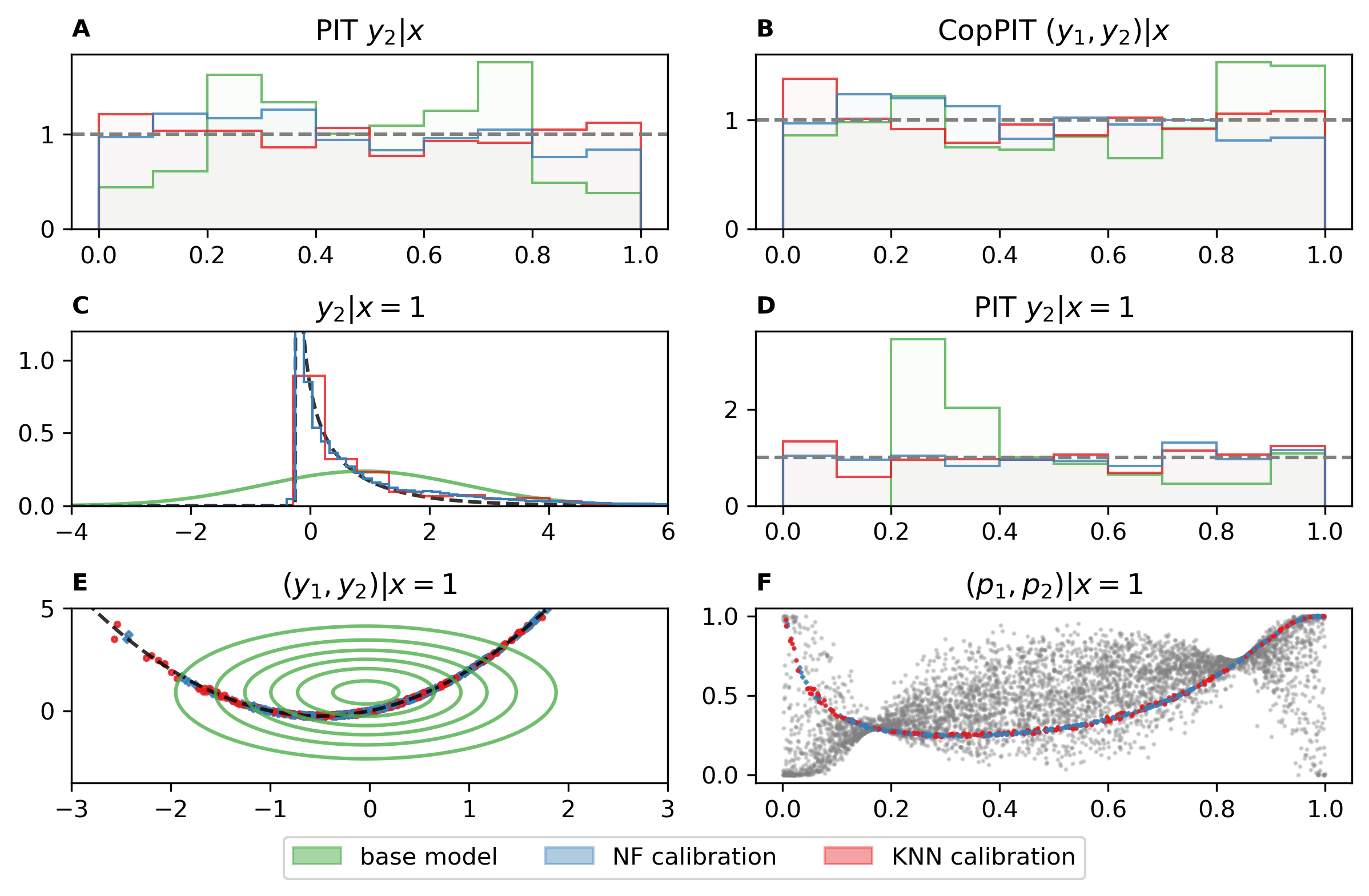}
    \caption{\small Simulation. Twisted Gaussians. \textbf{A), B)} PIT values for $Y_2\mid X$ and CopPIT values respectively for the base model (green), NF (blue) and KNN (red), calculated on a hold-out test set. \textbf{C)} Estimated densities for $p(y_2\mid x=1)$ under the three models (again indicated by color). The dashed black line gives the density under the true data generating process. \textbf{D)} PIT values for $Y_2\mid x=1$. \textbf{E)} Samples from the predictive distribution $p(y_1,y_2\mid x=1)$ under the NF model (blue diamonds) and the KNN model (red dots). For reference the contour plot from the base model is given in green. Under the true model all samples should lie on the dashed black line. \textbf{F)} Scatter plot of the PIT values from the validation set respective to the base model (grey). The PIT values corresponding to the samples shown in panel E are given in colour.}
    \label{fig:twistedGaussians}
\end{figure}

Figure~\ref{fig:twistedGaussians}A and Figure~\ref{fig:twistedGaussians}B show marginal PIT values for $Y_2\mid X$ and CopPIT values for $(Y_1,Y_2)\mid X$ respectively indicating that both the KNN and the NF approach result in multivariate calibrated models calculated on $n_\text{test}=1,000$ hold-out samples from the true DGP. However, our approach results not only in global calibration, but in local calibration in the following sense. Conditional on $x=1$ the base model for $Y_2\mid X$ is grossly misspecified as it assumes a Gaussian distribution while the true predictive distribution is heavily skewed and bounded by $-0.25$ from below (dashed black line in Figure~\ref{fig:twistedGaussians}C) and the recalibrated models match the shape of the true distribution. Under the NF approach an arbitrary large sample from the calibrated model can be generated, while the KNN approach is restricted to a fixed sample size depending on $n_\text{val}$. Figure~\ref{fig:twistedGaussians}E shows PIT values for $Y_2\mid x=1$ calculated from $\tilde{n}_\text{test}=1,000$ samples which are close to uniformity for both NF and KNN. The bivariate distribution $(Y_1,Y_2)\mid x=1$ is degenerate as all samples from the true model fulfill $Y_1^2+Y_1-Y_2=0$. Figure~\ref{fig:twistedGaussians}E shows samples from the calibrated models for $(Y_1,Y_2)\mid x=1$, which are virtually indistinguishable for NF and KNN and very close to the true distribution drastically improving the base model with independent marginals. Finally, Figure~\ref{fig:twistedGaussians}F shows how the joint marginal PIT values $\bm p^{(i)}$ from the base model under the validation set (shown in grey) encapsulate the complex dependence structure of the true model. The PIT values used by KNN and NF to generate the samples shown in panel E are marked by color. Note again that KNN is restricted by selecting PIT values from the validation set, which are sufficiently close to $x\approx 1$, while NF generates samples from an approximation to $(p_1,p_2)\mid x=1$, meaning that the approach could potentially hallucinate information not supported by the data, but allowing to draw an arbitrarily large sample from the recalibrated model.

\section{Additional Results for the Applications}\label{app:app}

\subsection{Currency exchange rates}

\begin{figure}[bt!]
    \centering
    \includegraphics[width=\textwidth,keepaspectratio]{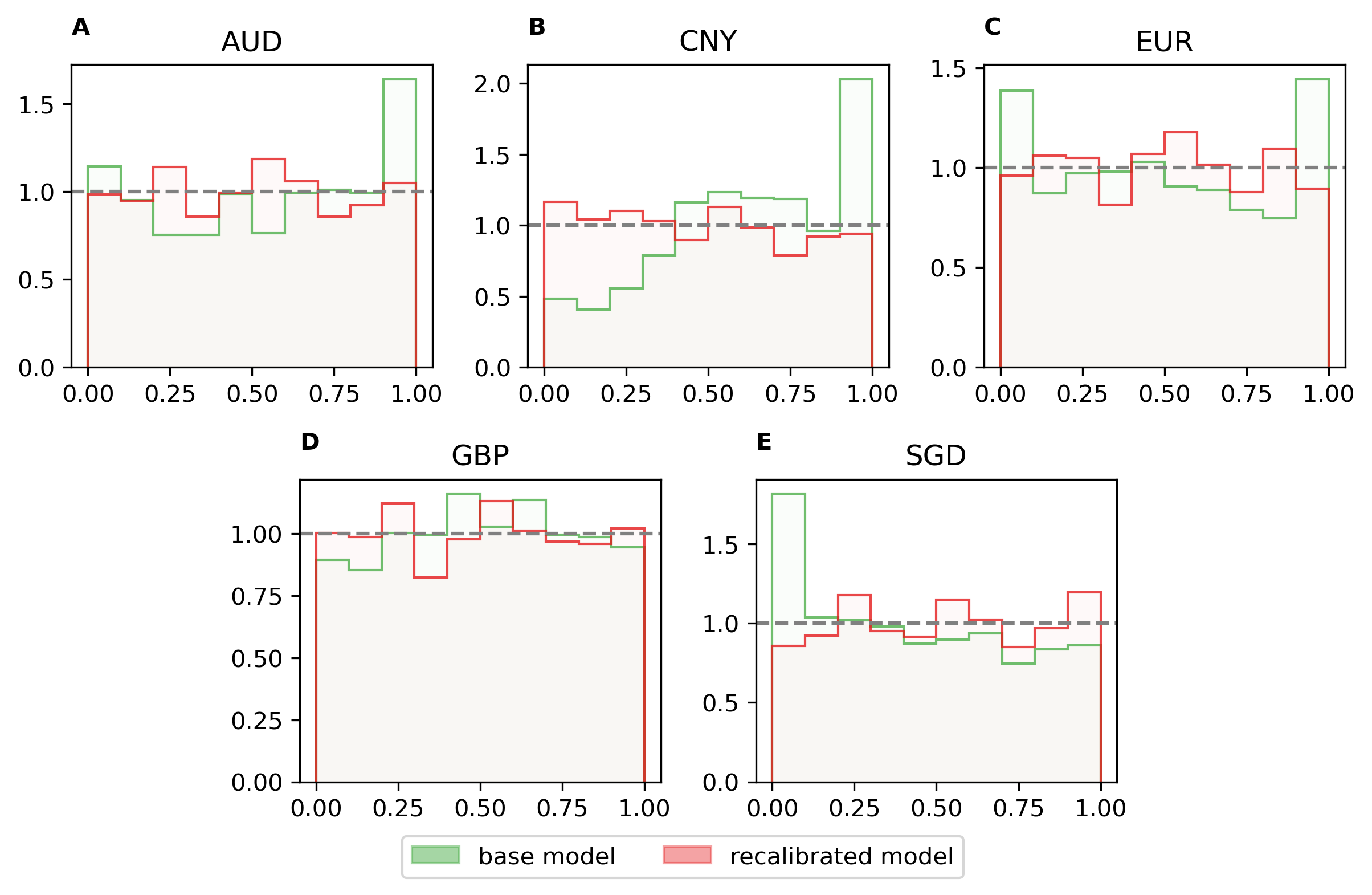}
    \caption{\small Currency exchange. Histograms of the univariate PIT values for the one-day ahead forecast for the five currencies \textbf{(A)}--\textbf{(E)} under the base model (green) and the recalibrated model (red).}
    \label{fig:cur1}
\end{figure}

Figure~\ref{fig:cur1} shows histograms of the univariate PIT values for the five currencies. Under the base model none of the margins is probability calibrated and the kind of miscalibration differs across currencies. Under the recalibrated model all margins are probability calibrated.

\subsection{Childhood malnutrition}

\begin{figure}[bt!]
    \centering
    \includegraphics[width=\textwidth,keepaspectratio]{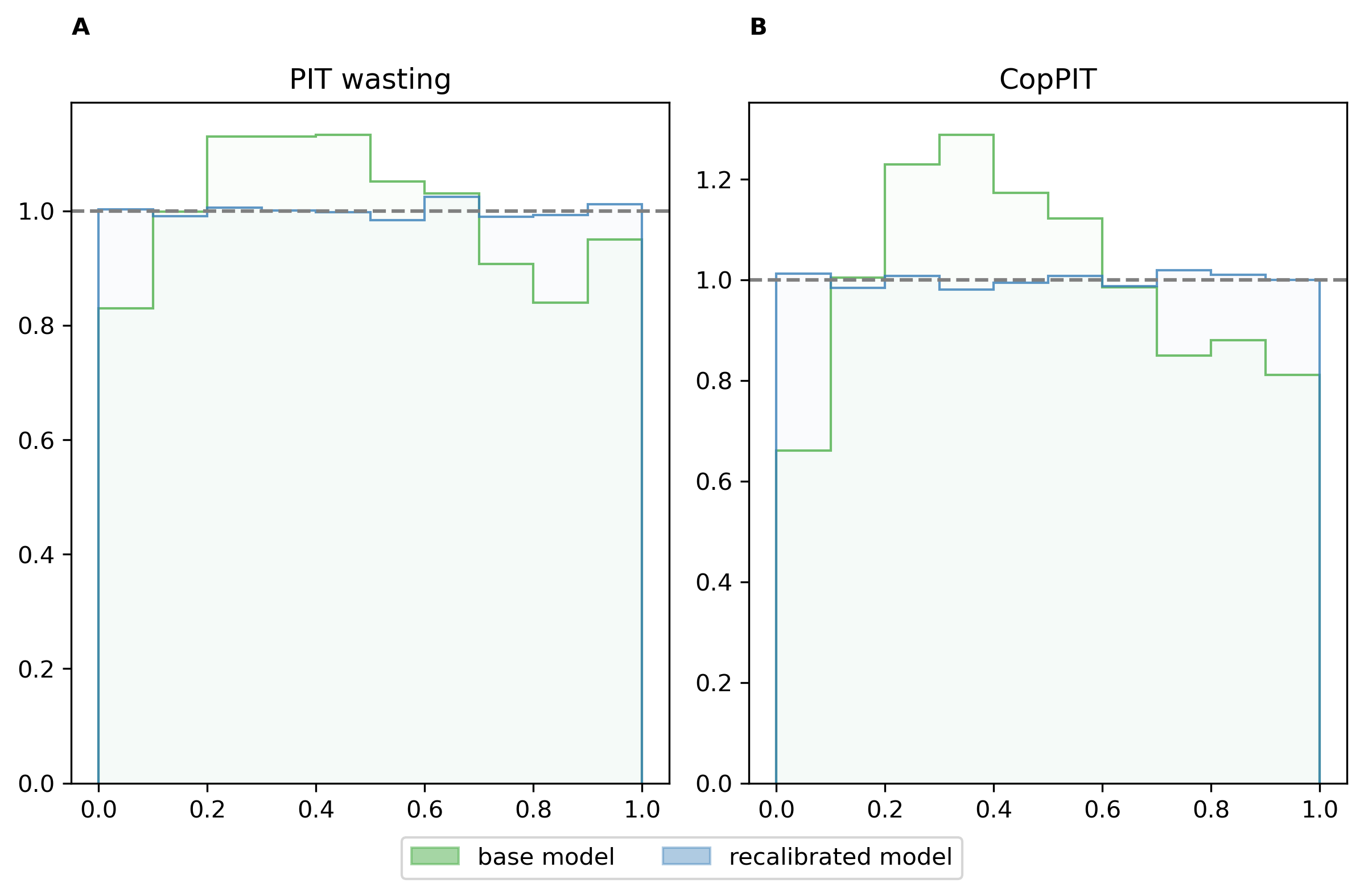}
    \caption{\small Malnutrition.  PIT values for \texttt{wasting} \textbf{(A)} and CopPIT values \textbf{(B)} respectively for the base model (green) and the recalibrated model (blue).}
    \label{fig:malnutrition_cal}
\end{figure}

\begin{figure}[bt!]
    \centering
    \includegraphics[width=\textwidth,keepaspectratio]{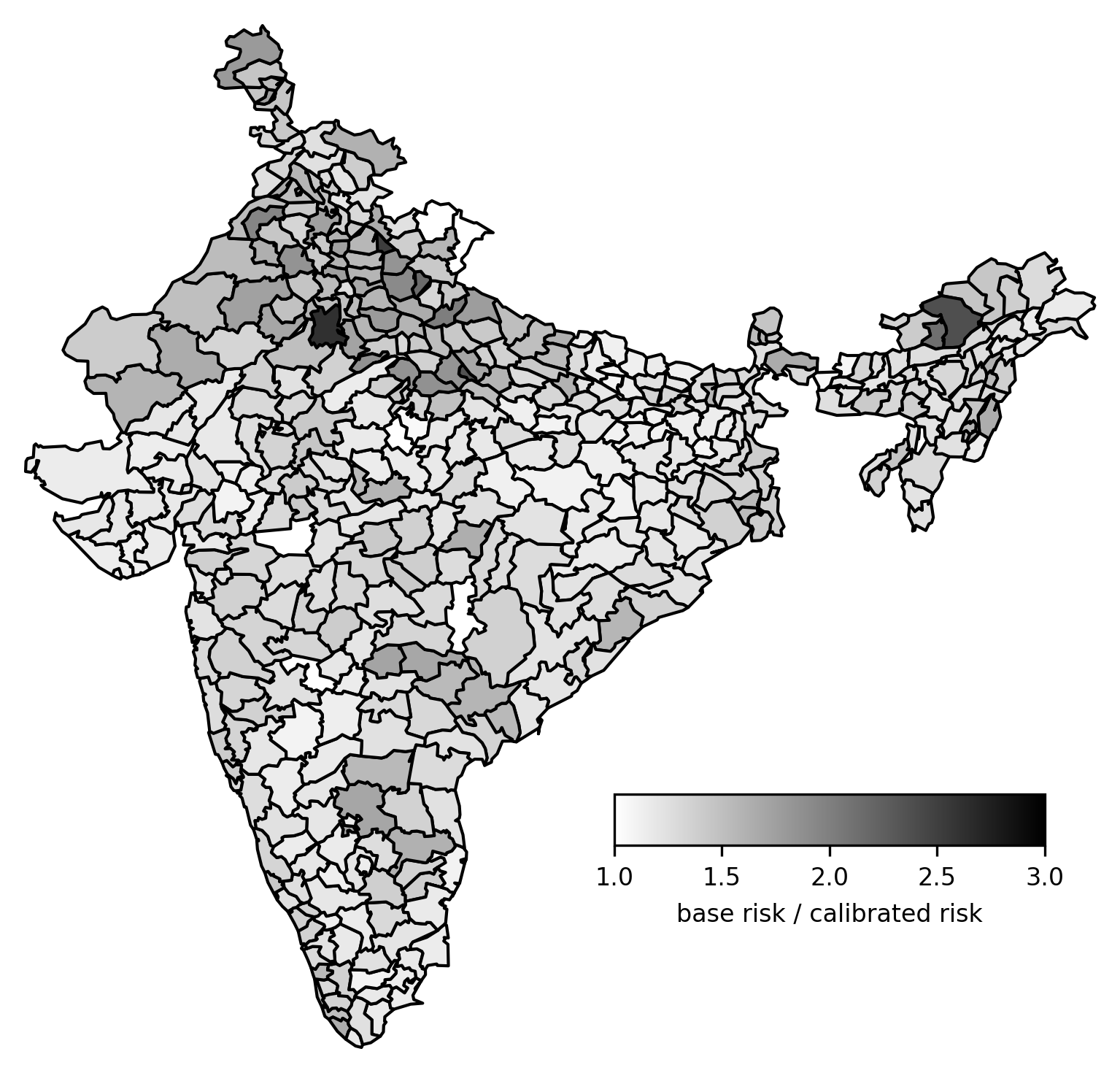}
    \caption{\small Malnutrition. Effect of the recalibration for the different districts. Shown is the fraction between the main effect for \texttt{dist} under the base model and the recalibrated model. Positive values indicate that the estimated risk is lower for the recalibrated model than in the base model and the relative magnitude of change is indicated by the shade of the region.}
    \label{fig:malnutrition2}
\end{figure}

Figure~\ref{fig:malnutrition_cal} shows histograms for the PIT and CopPIT values for the base and the recalibrated model indicating that NF improves the calibration of the regression model. Figure~\ref{fig:malnutrition2} shows how the main effect for \texttt{dist} differs between the base and the recalibrated model. 

\end{document}